%% file: paper1112.tex
\documentclass{llncs}
\usepackage{caption} 
\usepackage{amsmath,graphicx,wasysym}
\usepackage{hyperref}
\usepackage{bm}
\usepackage{booktabs}
\usepackage[colorinlistoftodos]{todonotes}
\usepackage{graphicx}
\usepackage{hyperref}

\begin{document}

\title{Evaluating 35 Methods to Generate Structural Connectomes Using Pairwise Classification} 

%
\author{Dmitry Petrov\inst{1}\inst{2} \and Alexander Ivanov\inst{2}\inst{4} \and
Joshua Faskowitz\inst{3} \and Boris Gutman\inst{1} \and \\ Daniel Moyer\inst{1} \and Julio Villalon\inst{1} \and Neda Jahanshad\inst{1} \and Paul Thompson\inst{1}}

%
\authorrunning{**} 
%
%

\institute{
Imaging Genetics Center, University of Southern California, Los Angeles, USA\\
  \email{to.dmitry.petrov@gmail.com}\\ 
 \and
 The Institute for Information Transmission Problems, Moscow, Russia
 \and 
 Indiana University, Bloomington, USA
 \and 
 Skoltech Institute of Science and Technology, Moscow, Russia\\
  }

\maketitle              

\begin{abstract}

There is no consensus on how to construct structural brain networks from diffusion MRI. How variations in pre-processing steps affect network reliability and its ability to distinguish subjects remains opaque. In this work, we address this issue by comparing 35 structural connectome-building pipelines. We vary diffusion reconstruction models, tractography algorithms and parcellations. Next, we classify structural connectome pairs as either belonging to the same individual or not. Connectome weights and eight topological derivative measures form our feature set. For experiments, we use three test-retest datasets from the Consortium for Reliability and Reproducibility (CoRR) comprised of a total of 105 individuals. We also compare pairwise classification results to a commonly used parametric test-retest measure, Intraclass Correlation Coefficient (ICC)\footnote{\small Code and results are available at \url{https://github.com/lodurality/35_methods_MICCAI_2017}}.

\end{abstract}
\section{Introduction}
\label{sec:intro}

\input{intro/intro.tex}

\begin{figure*}[t!]
\label{whole_pipe_fig}
\centering
\centerline{\includegraphics[width=13cm]{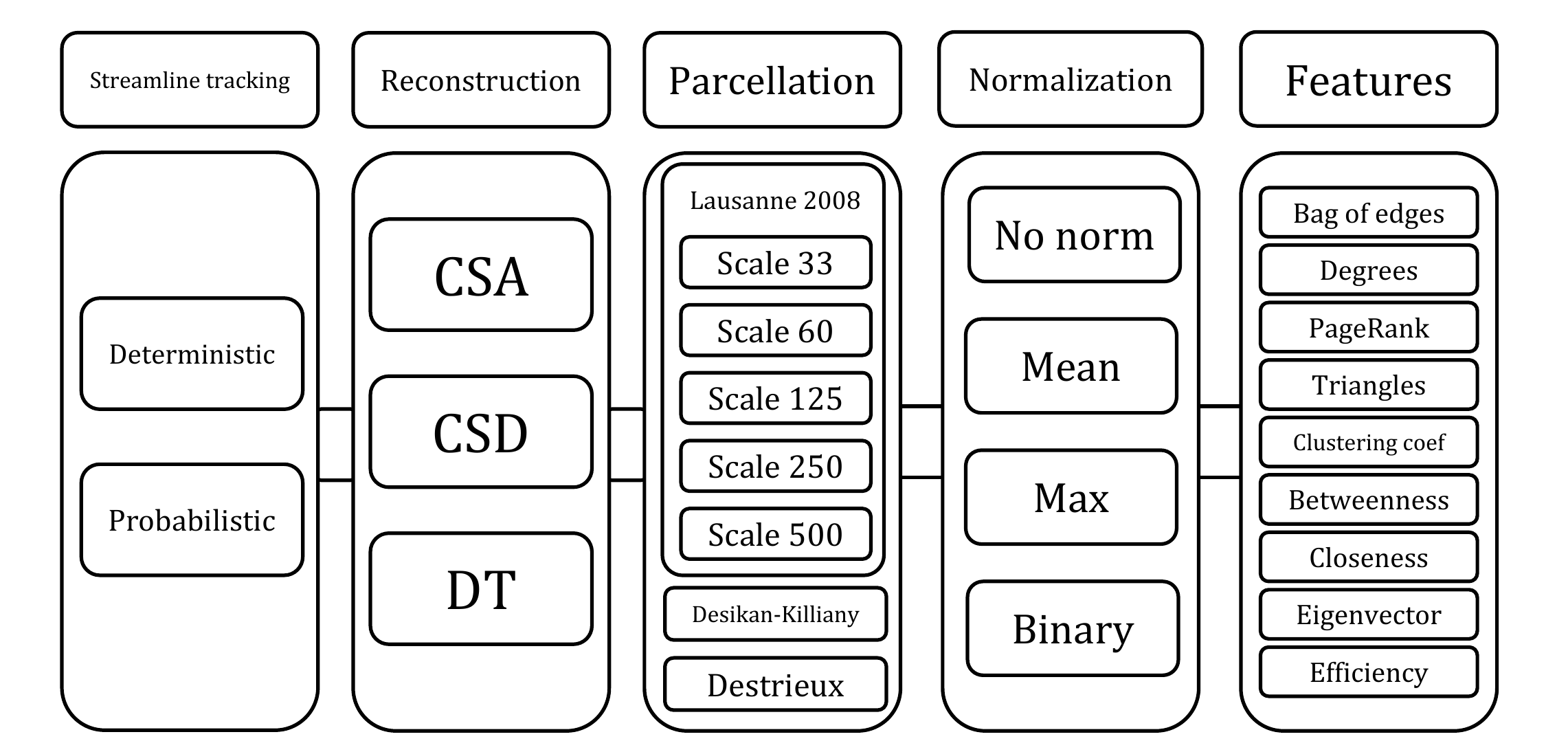}}
\caption{Overview of the parameter options for connectome construction and feature building pipelines. A complete description of each step can be found in sections \ref{features}, \ref{reconstruction}, and \ref{parcellations} respectively.}
\label{overview_fig}
\vspace{-0.75cm}
\end{figure*}

\input{pipe/pipe.tex}

\section{Experiments}

\subsection{Base data}

We used neuroimaging data from the Consortium for Reliability and Reproducibility (CoRR; \cite{zuo2014open}). Data sites within CoRR were chosen due to availability of T1-weighted (T1w) and diffusion-weighted images (DWI) with retest period less then two months. Full information about scanners and acquisition details is available on the CoRR website. T1-weighted images were parcellated using FreeSurfer 5.3 and the various atlases \cite{desikan2006automated} \cite{destrieux2010automatic} \cite{hagmann2008mapping}.
\vspace{-0.25cm}
\begin{table}
\centering
\begin{tabular}{ccccccccccc}
\toprule
\shortstack{Dataset \\ {}} & \shortstack{N \\ {}} & \shortstack{Scans per \\subject {}} & \shortstack{Age, \\ years} & \shortstack{Number \\ of females} & \shortstack{Retest period, \\ days} & \shortstack{DWI \\ directions {}} \\
\midrule
BNU 1   & 49 & 2 & 23.0 $\pm$ 2.3 & 23  & 33-55 & 30 \\
HNU 1   & 30 & 10 & 24.4 $\pm$ 2.4 & 15 &  3-40 & 30 \\
IPCAS 1 & 26 & 2 & 20.7 $\pm$ 1.7 & 19  &  5-29 & 60 \\
\end{tabular}
\caption{Information about datasets. N --- number of subjects.}
\vspace{-1.25cm}
\end{table}
\noindent 

\subsection{DWI preprocessing}

Diffusion weighted images (DWI) were corrected for head motion and eddy currents using FSL eddy\textunderscore correct with normalized mutual information. T1w images were aligned to the DWI in native space using FLS BBR \cite{greve2009boundary} and then used as a target for registration-based EPI artifact correction using a nonlinear ANTs SyN \cite{avants2011reproducible} warp constrained to the phase encoding axis of the images. DWI images were then rigidly aligned to the MNI152 space and interpolated linearly. Rotation of the b-vectors was performed accordingly for motion connection and linear alignment. 

Tractography was conducted in the MNI152 2mm isotropic space using the Dipy package \cite{garyfallidis2014dipy} (version 0.11). We used Dipy's LocalTracking module to generate probabilistic and deterministic streamline tractograms, using the aforementioned local models. The CSA and CSD models were computed using a spherical harmonics order of $6$. Streamlines were seeded in three random locations per white matter voxel, proceeded in 0.5 mm increments, and were retained if both ends terminated in voxels likely to be gray matter (based on partial volume estimation maps). All other streamline termination criteria were set to LocalTracking default parameters. Due to its single orientation nature, the Diffusion Tensor reconstruction model was not run with probabilistic tractography, leaving five of six possible Streamline Tracking/Reconstruction model combinations.

\subsection{Pairwise and sex classification}

For each set of connectomes described above we made all possible pairs of connectomes as described in \ref{pairs}. Using this technique we obtained 1176 pairs (49 of which were labeled as 1) from BNU 1 data, 44850 pairs from HNU 1 data (1350 of which were labeled as 1) and 325 pairs from IPCAS 1 data (26 of which were labeled as 1). Due to huge imbalance of classes in generated pairs, we used all samples with label 1 and equally sized random subsample of 0. Our result do not depend on a random state.

As an additional validation of our pipeline we perform sex classification on datasets using same combinations of connectome building steps, parcellations, normalizations and network features we used for pairwise classification. 

\section{Results}
\begin{figure}[!tbp]
\label{test}
  \centering
  \begin{minipage}[b]{0.49\textwidth}
    \includegraphics[width=\textwidth]{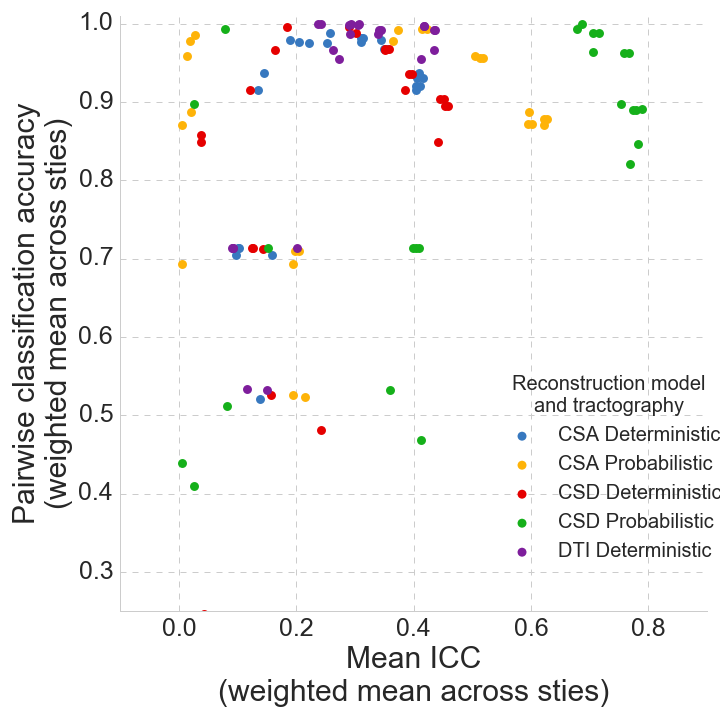}
  \end{minipage}
  \hfill
  \begin{minipage}[b]{0.49\textwidth}
    \includegraphics[width=\textwidth]{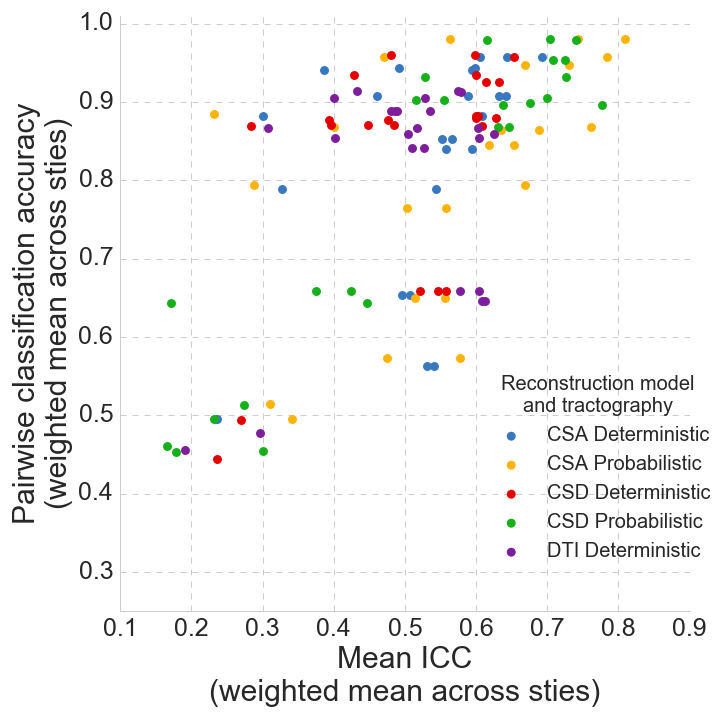}
  \end{minipage}
  \caption{Scatter plots for mean feature ICC and PACC for bag of edges (left) and closeness centrality (right) depending on reconstruction model, tractography, connectome normalization and parcellation. Each point represents a weighted mean of ICC/PACC across three datasets. ICC was weighted by the number of subjects and pairwise classification by the number of scans per subject.}
  \label{fig1}
  
\end{figure}

\figurename{\ref{fig1}} shows scatter plots of PACC vs ICC depending on the reconstruction model, tractography, normalization and parcellation. We see that the combination of CSA/CSD reconstruction model and probabilistic tractography performs best. Though excluded due to space limitations, weighted degrees, number of triangles, clustering coefficient and PageRank all have scatter patterns closely mimicking that for bag of edges; likewise, a pattern similar to closeness centrality holds for betweenness centrality, eigenvector centrality and local efficiency. 

\begin{figure}[!tbp]
  \centering
  \begin{minipage}[b]{0.49\textwidth}
    \includegraphics[width=\textwidth]{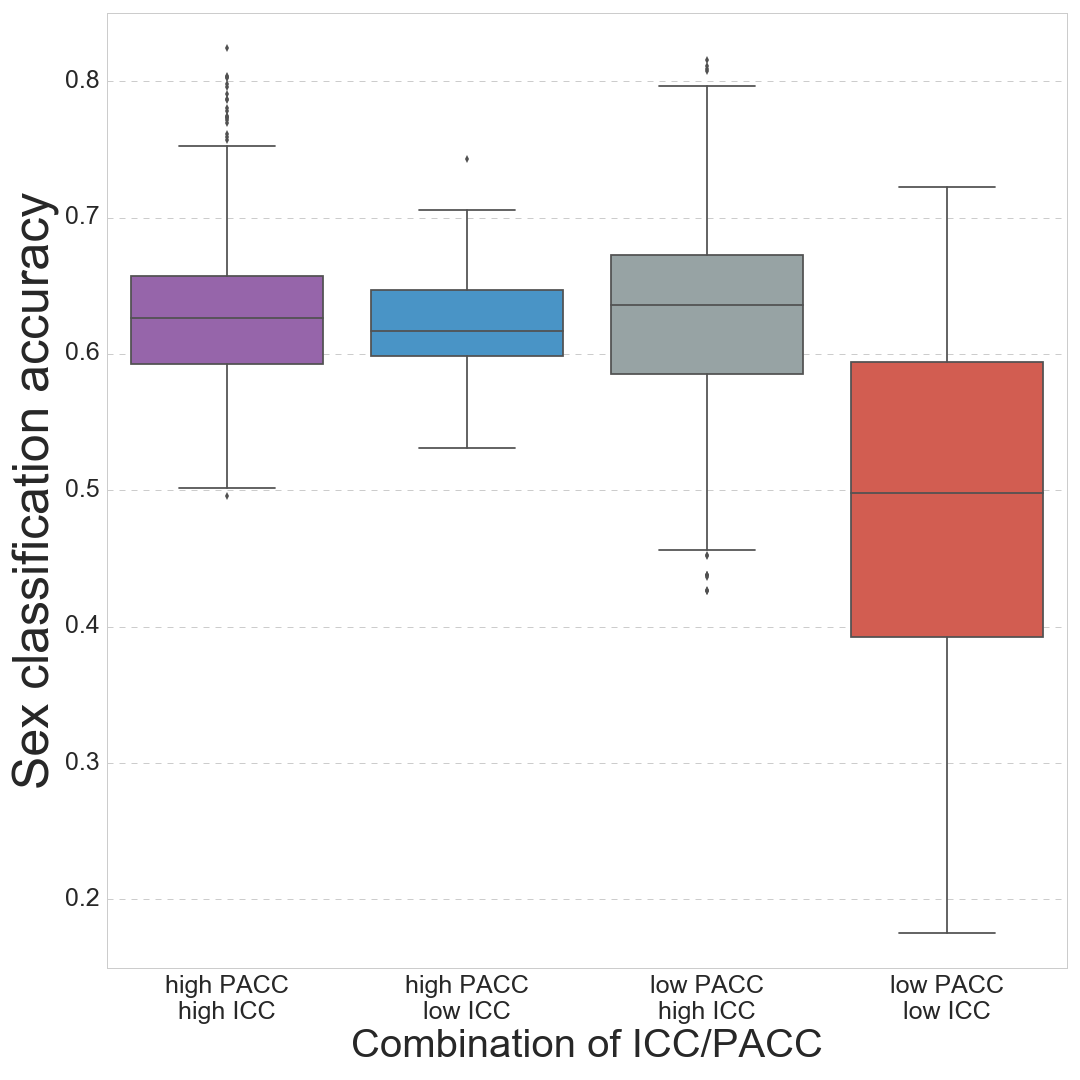}
    \end{minipage}
  \hfill
  \begin{minipage}[b]{0.49\textwidth}
    \includegraphics[width=\textwidth]{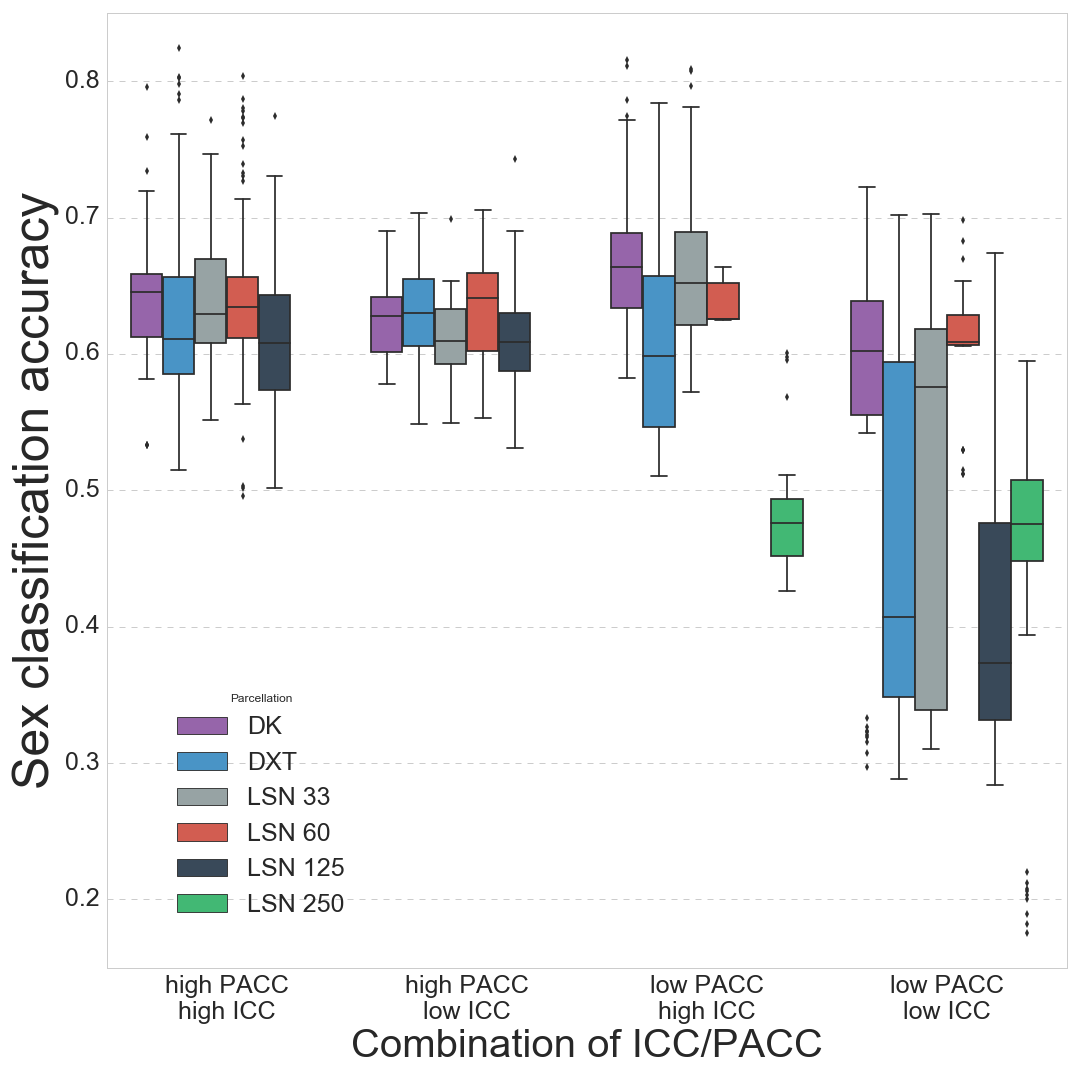}
    \end{minipage}
   \caption{Distribution of sex classification accuracy depending reconstruction model in terms of high/low pairwise classification accuracy and ICC (left), specified by parcellation (right). Each sex classification accuracy value is a weighted mean across datasets weighted by the number of subjects. Pairwise accuracy threshold was set at 0.9, ICC threshold was 0.6.}
     \label{fig2}
  
\end{figure}

\figurename{\ref{fig2}} shows the accuracy of the gender classification task in four groups: PACC $\ge$ 0.9 and ICC $\ge$ 0.6; PACC $<$ 0.9 and ICC $\ge$ 0.6; PACC $\ge$ 0.9 and ICC $<$ 0.6; PACC $<$ 0.9 and ICC $<$ 0.6. Unlike their  combination, neither low ICC nor low PACC alone was sufficient to predict poor accuracy.

\section{Conclusion}

In this paper, we presented a straight-forward method for evaluating brain connectivity construction pipelines from diffusion-weighted MRI, as well as their derived measures. Our method is a generalization of the traditional intraclass correlation coefficient, one based on pairwise classification. Our results so far suggest that the method may be useful in identifying overall trends in connectome usefulness beyond simply feautre-wise reliability measures, particularly with respect to DWI model choice and tractography approaches. As well, our results appear to confirm the intuition that having both low ICC and low pairwise classification accuracy generally leads to poor performance in unrelated classification tasks. It is also notable that PACC or ICC alone are not sufficient to identify reliably features poorly suited for our test classification task. Though the results are promising, they must be viewed with some skepticism given the limited nature of our validation. More data and more independent classification tasks for validation are required for more definitive rankings of network construction approaches and features in terms of their usefulness in neuroimaging studies.

%
%
\bibliographystyle{splncs03}
\bibliography{strings}


\end{document}

%% file: intro/intro.tex
In recent years, connectomics has become a popular form of analysis for neuroimaging data. The construction of structural connectomes, derived from diffusion MRI data, involves multiple pre-processing steps, each of which can be performed in a variety of different ways. It is often unclear, however, what the best combination of choices is, for specific dataset or application, or what their effect is on the resulting data. Due to the ongoing work on each of these steps, the number of possible processing pipelines is growing at a combinatorially remarkable rate.

There is also a very large number of graph summary statistics and derived features. While each has its purpose, merits, and historical derivation, it is again unclear which of these features capture the most relevant information for specific applications.

It is thus useful to investigate the effects of pre-processing choices on resulting connectivity models, and various information measures for their relative information content. It is difficult to assess each possible pipeline due to their number, but we believe it is important to narrow down the space of options. In this paper, we present reproducibility and simple classification task results for three diffusion models, with generated streamlines using two different methods, using seven different parcellations and four different normalization schemes, generating nine different graph features (\figurename{\ref{overview_fig}}). While it is unlikely there exists a single best pipeline for all data for all research objectives, in the present work we provide data on qualities we believe to be necessary, though not sufficient for inclusion in rigorous analyses, for the specific tasks we analyze.

As a basic sanity check for the usefulness of structural network generation methods and features, we propose to use pair-wise classification accuracy (PACC) as a multivariate potentially non-linear supplement to the usual intraclass correlation coefficient used in test-retest datasets. The task at hand, given a particular set of network features, is to separate network pairs arising from different scans of the same subject from pairs from different subjects. As additional validation, we evaluated accuracy in a sex classification task, using each of the network feature sets assessed before. 

Our main result is as follows: in terms of reproducibility and our simple classification tasks, probabilistic tracking using either a Constrained Spherical Deconvolution local model \cite{tournier2007robust}  or the Constant Solid Angle \cite{aganj2010reconstruction}  method gave the best combination of pairwise classification and mean ICC. This result is consistent across three datasets of healthy adult test-retest scans, with both a low and moderate number of diffusion angles (30 and 62). In terms of predicting poor performance in another classification task, the combination of low ICC and low pairwise classification accuracy appears to predict poor performance while either of these measures alone does not; this was also found consistently across datasets. Though we conjecture that such a "useless feature" identification would generalize to other classification tasks, this result is not sufficient to guarantee the generalization.

%% file: pipe/pipe.tex
\section{Structural Connectomics Pipelines}

We abstract the connectome construction process to the following steps (in order of processing): fitting a local diffusion model, reconstructing tracts, fitting a cortical parcellation and counting streamline endpoint pairs, normalization, and building connectome features (see \figurename{\ref{overview_fig}}). We first describe the options we assessed at each stage, then the methods by which we assessed them. For the remainder of the paper we denote a set of connectomes as $\{C^i_j\}$, where $j$ is an index of a subject and $i$ is an index of an image. 

\subsection{Reconstruction models and tractography}

\label{reconstruction}

We consider three widely-used methods for reconstruction  of white matter architecture \cite{daducci2014quantitative}. The \textbf{Diffusion Tensor} model (DTI) is by far the most well known, in that it is often incorrectly synonymous with diffusion-weighted imaging. DTI is also the simplest model, simply fitting an ellipsoidal diffusion pattern at each voxel. The \textbf{Constant Solid Angle} model (CSA) \cite{aganj2010reconstruction} produces orientation density functions that are generalizations of the ellipsoidal diffusion tensor to any continuous spherical function. This particular method is regularized and uses spherical harmonics to parameterize spherical functions. \textbf{Constrained Spherical Deconvolution} (CSD) \cite{tournier2007robust} models dominant tract directions as delta functions on the sphere convolved with a response kernel. The deconvolution recovers these directions from an estimated empirical ODF.

Once local models of fiber orientation have been constructed, a whole brain tractography reconstruction is applied. Here there are two general categories, \textbf{deterministic} tracking, which takes only the principle fiber direction, and \textbf{probabilistic}, which uses the full ODF and not simply the mode.

\subsection{Parcellations and Network Construction}
\label{parcellations}

There is a wide variety of parcellation choices. These have a non-trivial effect on the resulting graphs and derived graph measures \cite{van2010comparing}, and also, as we show here, on the consistency of those measures. We test the following parcellations, which were chosen based on their popularity and to represent a variety of scales: Desikan-Killiany (DK) \cite{desikan2006automated}, Destrieux \cite{destrieux2010automatic}, and the Lausanne 2008 (at five different scales) \cite{hagmann2008mapping}. We recorded the number of streamlines having endpoints in each pair of labels for each parcellation, using these counts as edge weights in each constructed connectome. The normalization of connectivity matrices also may be useful prior to any analysis (\cite{hagmann2007mapping}, \cite{bassett2011conserved}). We use the following three normalization schemes along with no normalization at all: mean, maximum, and binary normalization with zero threshold. 





\subsection{Network features}
\label{features}

For each connectome and each normalization we build ``bag of edges" vectors from the upper triangle of the adjacency matrix. In addition, we calculate eight network metrics for each node: \textbf{weighted degrees}, or strength; \textbf{closeness}, \textbf{betweenness} and \textbf{eigenvector centralities}; \textbf{local efficiency}; \textbf{clustering coefficient}; and \textbf{weighted number of triangles} around node. We choose these features because they are well-described and reflect different structural properties of connectomes \cite{rubinov2010complex}. We also calculate PageRank for each node. Introduced in 1998 by Brin and Page \cite{pagerank_original} this metric roughly estimates probability that a random walk on the network will be observed at particular node.

\subsection{Pairwise features}
\label{pairs}

Each normalization and set of features described above defines a mapping from connectome space to feature space $C \rightarrow f(C)$. As our goal is to check how well this mapping separates connectomes in it, we propose various pairwise features. For each set of connectome features in question we make all possible pairs of connectome features -- ($f(C^{i_1}_{j_1}), f(C^{i_2}_{j_2})$). For each pair, we assign a binary target variable -- 1 if connectomes are from the same subject ($j_1 = j_2$), 0 -- if they are from different subjects ($j_1 \neq j_2$). Finally, for each pair we build a vector of three features, describing their difference $\|f(C_1) - f(C_2) \|$ according to $l_1$, $l_2$ and $l_{\infty}$ norms.

\subsection{Classification models and validation}

We use linear classifiers for pairwise  and sex classification problems: logistic regression, SVM with linear kernel and stochastic gradient descent (SGD) with modified Huber loss. We scale features with standard scaling and apply elastic-net regularization for each of the classifiers.

We measure model performance and accuracy, in a two-step validation procedure. First, for each dataset, we perform hyperparameter grid search based on a 5-fold cross-validation with a fixed random state for reproducibility. For each model, we varied the overall regularization parameter, $l_1$-ratio and number of iterations for SGD. Then we evaluate the best parameters on 50 train/test splits with fixed different random states (test size was set to 20\% of data). We characterize each connectome building pipeline and feature by mean pairwise/sex classification accuracy on these 50 test splits.

\subsection{Reproducibility measure}

As a reproducibility measure for connectome mapping, we use the Intraclass Correlation Coefficient \cite{shrout1979intraclass} between measurements taken at different time points: 
$$
ICC = \frac{BMS-WMS}{BMS + (k-1)WMS},
$$
where BMS is the between-subject mean sum of squares, WMS is the within-subject mean sum of squares, and $k$ is the number of scans per subject. For each pipeline and derived set of features we exclude features with no variation and calculate mean ICC value for remaining features, thus characterizing this pipeline by one ICC value.

%% file: paper1112.bbl
\begin{thebibliography}{10}
\providecommand{\url}[1]{\texttt{#1}}
\providecommand{\urlprefix}{URL }

\bibitem{aganj2010reconstruction}
Aganj, I., Lenglet, C., Sapiro, G., Yacoub, E., Ugurbil, K., Harel, N.:
  Reconstruction of the orientation distribution function in single-and
  multiple-shell q-ball imaging within constant solid angle. Magnetic Resonance
  in Medicine  64(2),  554--566 (2010)

\bibitem{avants2011reproducible}
Avants, B.B., Tustison, N.J., Song, G., Cook, P.A., Klein, A., Gee, J.C.: A
  reproducible evaluation of {ANTS} similarity metric performance in brain
  image registration. Neuroimage  54(3),  2033--2044 (2011)

\bibitem{bassett2011conserved}
Bassett, D.S., Brown, J.A., Deshpande, V., Carlson, J.M., Grafton, S.T.:
  Conserved and variable architecture of human white matter connectivity.
  Neuroimage  54(2),  1262--1279 (2011)

\bibitem{daducci2014quantitative}
Daducci, A., Canales-Rodr{\i}, E.J., Descoteaux, M., Garyfallidis, E., Gur, Y.,
  Lin, Y.C., Mani, M., Merlet, S., Paquette, M., Ramirez-Manzanares, A.,
  et~al.: Quantitative comparison of reconstruction methods for intra-voxel
  fiber recovery from diffusion mri. IEEE transactions on medical imaging
  33(2),  384--399 (2014)

\bibitem{desikan2006automated}
Desikan, R.S., S{\'e}gonne, F., Fischl, B., Quinn, B.T., Dickerson, B.C.,
  Blacker, D., Buckner, R.L., Dale, A.M., Maguire, R.P., Hyman, B.T., et~al.:
  An automated labeling system for subdividing the human cerebral cortex on mri
  scans into gyral based regions of interest. Neuroimage  31(3),  968--980
  (2006)

\bibitem{destrieux2010automatic}
Destrieux, C., Fischl, B., Dale, A., Halgren, E.: Automatic parcellation of
  human cortical gyri and sulci using standard anatomical nomenclature.
  Neuroimage  53(1),  1--15 (2010)

\bibitem{garyfallidis2014dipy}
Garyfallidis, E., Brett, M., Amirbekian, B., Rokem, A., Van Der~Walt, S.,
  Descoteaux, M., Nimmo-Smith, I.: Dipy, a library for the analysis of
  diffusion mri data. Frontiers in neuroinformatics  8, ~8 (2014)

\bibitem{greve2009boundary}
Greve, D., Fischl, B.: A boundary-based cost function for within-subject,
  cross-modal registration. Neuroimage  47,  S100 (2009)

\bibitem{hagmann2008mapping}
Hagmann, P., Cammoun, L., Gigandet, X., Meuli, R., Honey, C.J., Wedeen, V.J.,
  Sporns, O.: Mapping the structural core of human cerebral cortex. PLoS Biol
  6(7),  e159 (2008)

\bibitem{hagmann2007mapping}
Hagmann, P., Kurant, M., Gigandet, X., Thiran, P., Wedeen, V.J., Meuli, R.,
  Thiran, J.P.: Mapping human whole-brain structural networks with diffusion
  mri. PloS one  2(7),  e597 (2007)

\bibitem{pagerank_original}
Page, L., Brin, S., Motwani, R., Winograd, T.: Tech. Rep. 1999-66 (November)

\bibitem{rubinov2010complex}
Rubinov, M., Sporns, O.: Complex network measures of brain connectivity: uses
  and interpretations. Neuroimage  52(3),  1059--1069 (2010)

\bibitem{shrout1979intraclass}
Shrout, P.E., Fleiss, J.L.: Intraclass correlations: uses in assessing rater
  reliability. Psychological bulletin  86(2),  420 (1979)

\bibitem{tournier2007robust}
Tournier, J.D., Calamante, F., Connelly, A.: Robust determination of the fibre
  orientation distribution in diffusion mri: non-negativity constrained
  super-resolved spherical deconvolution. NeuroImage  35(4),  1459--1472 (2007)

\bibitem{van2010comparing}
Van~Wijk, B.C., Stam, C.J., Daffertshofer, A.: Comparing brain networks of
  different size and connectivity density using graph theory. PloS one  5(10),
  e13701 (2010)

\bibitem{zuo2014open}
Zuo, X.N., Anderson, J.S., Bellec, P., Birn, R.M., Biswal, B.B., Blautzik, J.,
  Breitner, J.C., Buckner, R.L., Calhoun, V.D., Castellanos, F.X., et~al.: An
  open science resource for establishing reliability and reproducibility in
  functional connectomics. Scientific data  1 (2014)

\end{thebibliography}
